\documentclass{article}

\usepackage[nonatbib]{nips_2017}
\usepackage{nips_2017} 
\usepackage{caption}
\usepackage{subcaption}
\usepackage{cite}
\usepackage[utf8]{inputenc} 
\usepackage[T1]{fontenc}    
\usepackage{url}            
\usepackage{graphicx}
\usepackage{booktabs}       
\usepackage{amsfonts}       
\usepackage{nicefrac}       
\usepackage{microtype}      
\usepackage{amsmath}
\usepackage{float}
\usepackage{amssymb}
\usepackage[bookmarks=false]{hyperref}

\usepackage{lineno}

\title{Learning graph-Fourier spectra of textured surface images for defect localization}

\author{ff}

\begin{document}

\maketitle

\begin{abstract}
In the realm of industrial manufacturing, product inspection remains a significant bottleneck, with only a small fraction of manufactured items undergoing inspection for surface defects. Advances in imaging systems and AI can allow automated full inspection of manufactured surfaces. However, even the most contemporary imaging and machine learning methods perform poorly for detecting defects in images with highly textured backgrounds, that stem from diverse manufacturing processes. This paper introduces an approach based on graph Fourier analysis to automatically identify defective images, as well as crucial graph Fourier coefficients that inform the defects in images amidst highly textured backgrounds. The approach capitalizes on the ability of graph representations to capture the complex dynamics inherent in high-dimensional data, preserving crucial locality properties in a lower-dimensional space. A convolutional neural network model (1D-CNN)  was trained with the coefficients of the graph Fourier transform of the images as the input to identify, with classification accuracy of 99.4\%, if the image contains a defect. An explainable AI method using SHAP (SHapley Additive exPlanations) was used to further analyze the trained 1D-CNN model to discern important spectral coefficients for each image. This approach sheds light on the crucial contribution of low-frequency graph eigen waveforms to precisely localize surface defects in images, thereby advancing the realization of zero-defect manufacturing.
\end{abstract}

\textbf{Keywords:} Quality assurance; Surface defect localization; Graph fourier transforms; Convolutional neural networks; Explainbale AI

\section{Introduction}
In recent years, the field of quality control in manufacturing has witnessed a significant transformation owing to advancements in imaging systems. A study by GlobeNewswire, projects a remarkable surge for AI-based visual systems in manufacturing, anticipating an estimated global market value of \$21.3 billion by 2028, with a compelling compound annual growth rate of 43.4\% \cite{GNW}. The augmented availability and heightened sophistication of AI-based vision systems have transcended their role beyond traditional end-of-line product inspection to encompass on-line, intermittent and in-situ quality monitoring \cite{arinez2020artificial, konstantindis2018vision}. Imaging systems have evolved into indispensable tools for identifying a myriad of morphological and geometric defects, including scratches, holes, warping, and inclusions, throughout various stages of the manufacturing process \cite{park2016machine}.

\begin{figure}[ht]
    \centering
    \includegraphics[width=0.75\linewidth]{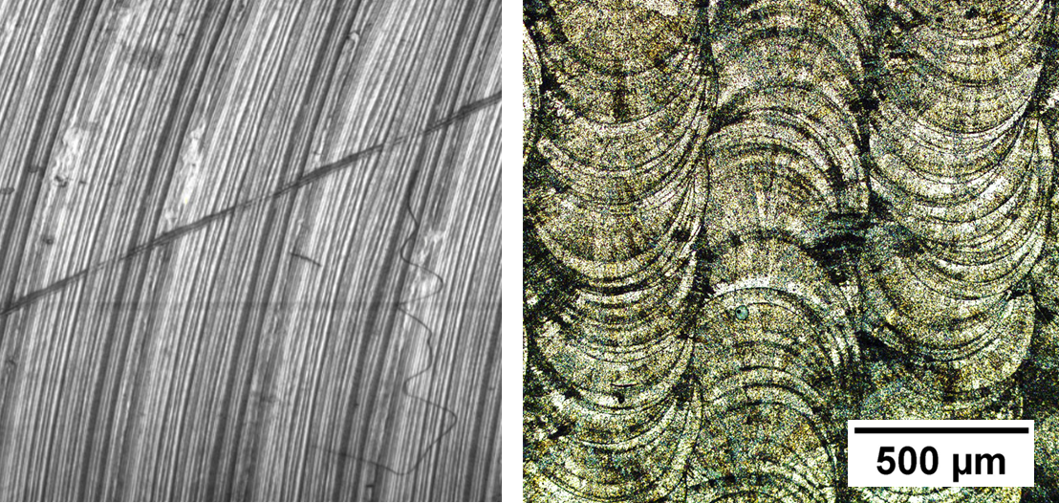}
    \captionsetup{font=normal}
    \caption{Images illustrating distinct complex surface textures resulting from a turning process (left) and ripple formation from laser direct energy deposition (L-DED) process (right) \cite{balhara2023ripple} }
    \label{Fig:Figure 1}
\end{figure}

One key aspect that contributes to the complexity of defect identification is the diverse range of textures generated by different manufacturing processes. In metallic manufacturing, casting processes such as sand casting or investment casting yield textures characterized by irregularities stemming from the cooling and solidification of molten metal. In additive manufacturing, textures emerge as a result of layer-wise deposition of material, frequently accompanied by porosity and shrinkage defects \cite{akhil2020image, balhara2023ripple, triantaphyllou2015surface}. Machining processes, including milling and turning, impart distinct surface finishes with tool marks and cutting patterns, contributing to the overall texture. Turning or cutting operations performed on a lathe often lead to defects like burrs, grooves, adhered material, and tearings \cite{bordin2014effect, ulutan2011machining}. Plastic and polymer manufacturing, specifically injection molding, introduces textures contingent on mold surface finish and polymer type, with defects like sink marks and weld lines influencing the visual appearance. In the realm of composite manufacturing, lay-up processes shape textures through the arrangement of fibers and resin distribution, while compression molding imparts textures affected by molding surface characteristics. Semiconductor manufacturing involves intricate processes yielding specific surface textures influenced by factors such as etching and deposition techniques, and defects in these processes can profoundly impact the chip's functionality \cite{nakazawa2019anomaly, saqlain2020deep}. 
In contrast to open-source datasets such as MVTEC-AD \cite{bergmann2019mvtec} that primarily focus on defect localization, machining processes generate intricate background textures, as depicted in Figure \ref{Fig:Figure 1}. The complexity of these textures underscores the significance of understanding their diversity, representing a crucial aspect in advancing methodologies for defect identification within manufacturing contexts.

From an imaging perspective, defects emerge as irregularities embedded within the diverse textures, presenting a myriad of forms that pose challenges for precise classification and localization. The intricacies reside on the subtle interactions of pixel intensities within textures, and their accurate identification holds key to unveiling the underlying defect formation mechanisms across diverse manufacturing processes.  Nonetheless, this precision remains elusive with current image segmentation methods. Traditional supervised learning methods \cite{akhil2020image,caggiano2019machine, park2016machine, saqlain2020deep}, while effective, often demand substantial amounts of labeled data for each type of defect. Despite their prowess in defect classification, these methods encounter limitations in localizing defects accurately and inferring the underlying defect formation mechanisms. Additionally, the need for a diverse dataset that spans the range of possible defects across manufacturing processes poses a practical challenge. Ganatma et al. \cite{nakkina2022smart} developed an AI Toolbox to relate distinct class of defects with specific image processing methodologies such as Wavelet Analysis, Morphological Component Analysis (MCA), and Basic Line Detector (BLD). They proposed an empirical recommendation formula based on three image metrics$-$entropy, Universal Quality Index (UQI), and Rosenberger's$-$to evaluate the performance of a method across a specified class of images.

As an alternative paradigm, graph signal processing (GSP) methods aim to leverage the relationship between signals and the topology of the graph where they are supported, to make meaningful inferences \cite{dong2020graph, ortega2018graph,shuman2013emerging}. The intuition behind using graph representations and analysis can be associated to the approach of Belkin and Niyogi \cite{belkin2003laplacian}, where they highlighted that the graph Laplacian can capture the complex dynamics of high-dimensional data and locality preserving properties in a low-dimension space, namely, the graph $G(V, E)$. Algebraic graph theoretic approaches have been adopted for process monitoring and anomaly detection in ultraprecision machining using multivariate time series sensor data \cite{montazeri2018sensor, tootooni2016spectral, rao2015quantification}. Bukkapatnam et al. \cite{bukkapatnam2018planar} employed a planar random graph representation to monitor the surface morphology evolution of electron beam printed Ti-6Al-4V samples during polishing and established an endpoint criterion based on Fiedler number $\lambda_2$, that is sensitive to the neighbourhood asperity structure. 

Spectral graph theoretic approaches have been previously used to describe topological relationships in various physical domains involving image processing \cite{cheung2018graph, iquebal2020consistent, shi2000normalized}. Hunag et al. \cite{huang2018graph} utilized GSP framework to analyze brain activity aligned with the structural brain graph, revealing signal variability during attention-switching tasks.They emphasized the framework's versatility for conducting analyses on both functional and structural connectivity in functional imaging datasets. However, in most real world scenarios, the underlying graph topology is not readily available, and needs to be inferred from the signals itself. The construction of a meaningful graph topology from the data is crucial for effective representation and inferring functional connectivity within the graph \cite{dong2019learning, mateos2019connecting}. Models for developing graph topology $G$ from data are either defined based on the underlying physical phenomenon on the graph like heat diffusion, or statistically modelled as a function that draws realization from a probability distribution over the variables that are representative of the graph structure \cite{koller2009probabilistic}.

In this study, we harness the principles of Graph Fourier Transform (GFT) to effectively localize the presence of surface defects in images. Utilizing GFT introduces a powerful means of handling complex spatial relationships within images with textured backgrounds. By conceptualizing the image as a signal distributed across a graph, where pixels are interconnected based on their spatial relationships, we aim to uncover specific graph spectral coefficients crucial for distinguishing and localizing defects. The graph spectral coefficients pave way to a more robust identification of nuanced morphological characteristics, contributing to superior accuracy in localizing defects amidst intricate image structures. A convolutional neural network (1D-CNN) classifier is employed to categorize images with diverse surface textures, discerning between those devoid of defects and those exhibiting defects with varying shapes, sizes, and locations, encompassing a spectrum of morphological characteristics. Inferences from subsequent explainable AI method using SHAP (SHapley Additive exPlanations), helps identify important spectral coefficients for defect localization within each image, and forms the basis for establishing correlations with the underlying physical mechanisms responsible for defect formation. This graph-based approach enables us to capture both global patterns and localized variations, making it a powerful tool for precise defect identification—an essential element in achieving zero-defect manufacturing. 

The rest of this paper is organized as follows: Section 2 provides a comprehensive overview of graph signal processing and its application to surface defect localization. In Section 3, we delineate the architecture of the 1D-CNN model and evaluate its performance. In Section 4, we delve into the interpretability of our model's predictions through SHAP analysis to infer how spectral coefficients contribute to defect localization, followed by conclusions in Section 5.

\section{Spectral graph theoretic approach for defect localization}

We introduce the formulation of graph and graph signals, i.e., signals whose samples are indexed by the nodes of an arbitrary graph. Let us define some of the notations associated with graph signals, graph fourier transform and the interpretation of graph variation operators. 

Consider a weighted graph $G = (V, E, W)$ where $V = \{1, 2, \ldots, N\}$ is a finite set of $N$ nodes or vertices, $E \in V \times V$ is a set of edges defined as ordered pairs $(i, j)$ representing pairwise relations between vertices, and $W: E \rightarrow \mathbb{R}$ is a mapping from the set of edges to scalar values $w_{ij}$ representing the edge weight between vertex $i$ to vertex $j$. A graph signal $X:V \rightarrow \mathbb{C}$ is a mapping from the vertices of the graph to complex numbers $X \in \mathbb{C}^N$. This graph signal can be represented as a vector $X=\{x_1, x_2, x_3, \ldots, x_N\}$ where $x_k$ indicates the value of the signal at the $k^{th}$ vertex in $V$.

\begin{figure*}[ht]
\centerline{\includegraphics[width=1.08\linewidth]{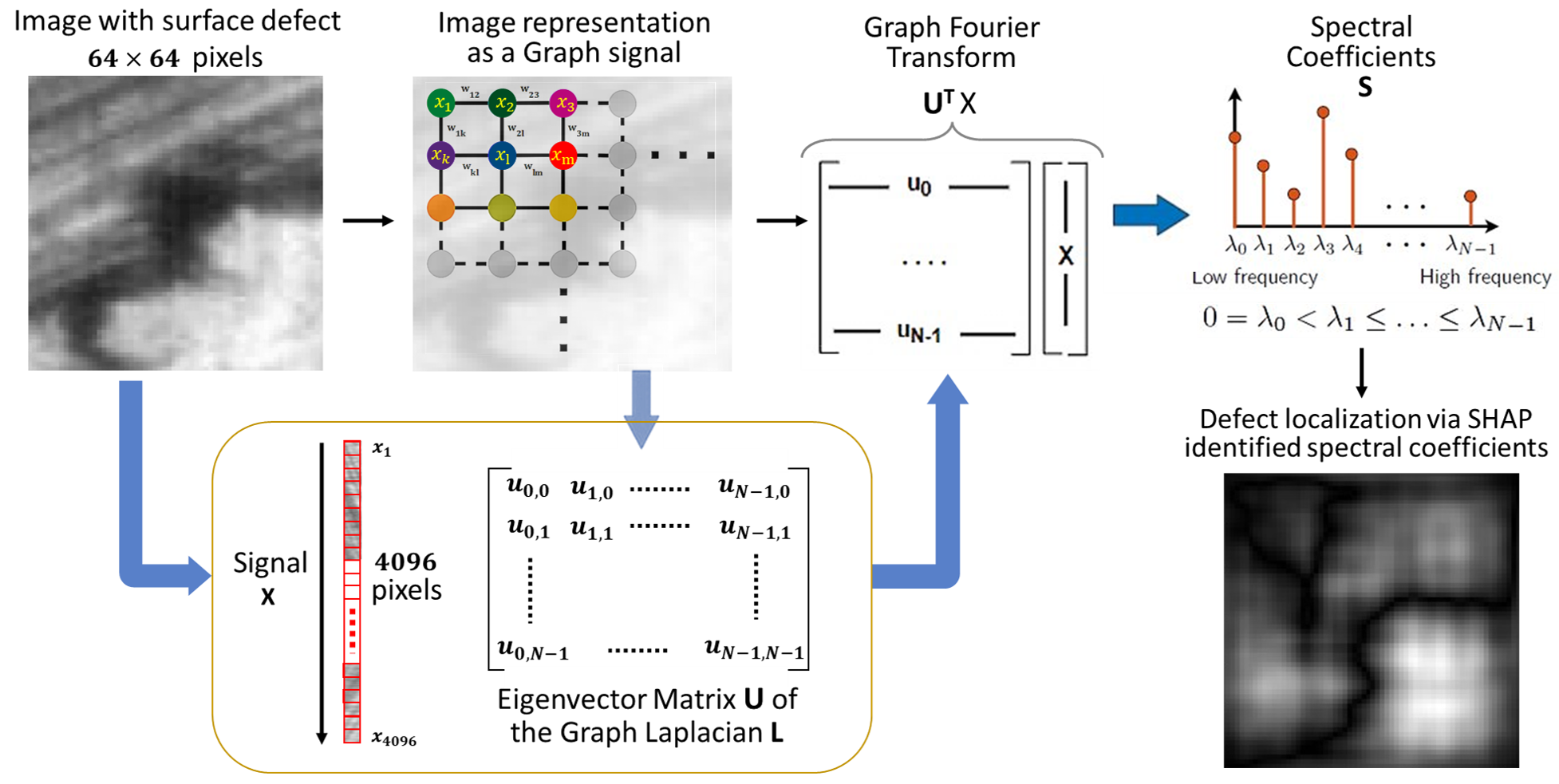}}
\captionsetup{font=normal}
\caption{Schematic representation of Graph Fourier Transforms for spectral coefficients extraction and defect localization}
\label{Fig: Figure 2}
\end{figure*} 

Here, we define a grid-based undirected graph structure on an image (see Figure \ref{Fig: Figure 2}) by representing each pixel as a node and connecting adjacent pixels with edges $(E)$ defined as ordered pairs $(i,j)$ where $(i)$ and $(j)$ are pixel indices. In contrast to traditional image segmentation methods, where edge weights are determined based on pixel intensities utilizing exponential or gaussian functions, our approach formulates a graph structure solely based on spatial relationships, specifically considering immediate neighbors in the $x$ and $y$ directions. The associated adjacency matrix $A \in \mathbb{R}^{N \times N}$ for the graph is defined as

\begin{equation}
    A_{ij} = 
    \begin{cases}
        w_{ij} = 1, & \text{if } (i, j) \in E \\
        0, & \text{otherwise}
    \end{cases}
\end{equation}

\noindent The degree matrix $D \in \mathbb{R}^{N \times N}$ for the graph is a diagonal matrix defined as follows:

\begin{equation}
    D_{ii} = \text{deg}(i) = \sum_{j \in n(i)} w_{ij} 
\end{equation}

\noindent where $n(i)$ stands for the neighborhood of vertex $i$. The laplacian matrix $L \in \mathbb{R}^{N \times N}$ for the above defined graph structure $G$ with adjacency matrix $A$ and degree matrix $D$ is defined as  

\begin{equation}
    L = D - A
\end{equation}

\noindent For undirected graphs with real and non-negative edge weights, the graph Laplacian matrix $L$ inherits properties of being real, symmetric, and positive semi-definite \cite{hu2021graph}. 

\subsection{Eigen decomposition of the graph laplacian matrix}
The real and symmetric nature of the graph Laplacian matrix \(L\) allows for eigen decomposition into a complete set of orthonormal eigenvectors \(u_0, u_1, \ldots, u_{N-1}\) and corresponding eigenvalues, in ascending order \(0 = \lambda_0 < \lambda_1 \leq \ldots \leq \lambda_{N-1}\):

\begin{equation}
    L = U \Lambda U^T 
\end{equation}

\noindent where \(U\) is the eigenvector matrix containing the eigenvectors \(u_0, u_1, \ldots, u_{N-1}\) as columns, satisfying \(U^{-1} = U^T\), and \(\Lambda\) is a diagonal matrix containing the eigenvalues, i.e., \(\Lambda = \text{diag}(\lambda_0, \lambda_1, \ldots, \lambda_{N-1})\). Each eigenvector \(u_k\) of the graph laplacian matrix \(L\) satisfies \(L u_k = \lambda_k u_k\). The eigenvalue \(\lambda_k\) is referred to as the graph frequency/spectrum, with a smaller eigenvalue corresponding to a lower graph frequency \cite{hu2021graph}.

The eigen decomposition of the graph Laplacian matrix \(L\) into eigenvectors \(U\) and corresponding eigenvalues \(\Lambda\) embodies a profound depiction of the graph's underlying structure and dynamics. In the context of a grid structure representing an image, where each edge directly connects neighboring pixels and bears a unit weight, the derived eigenvectors \(U\) offer nuanced insights into the image's representation. The eigenvector linked to \(\lambda_0\) acts as a fundamental mode, resembling a constant function and offering a comprehensive perspective on the overall structure of the image grid. Similarly, the eigenvector associated with \(\lambda_1\) (second smallest eigen value), widely recognized as the Fiedler vector, encapsulates vital information regarding the graph's connectivity, acting as a representation of the first non-trivial mode.

\begin{figure*}[ht]\vspace*{4pt}
\centerline{\includegraphics[width=1.08\linewidth]{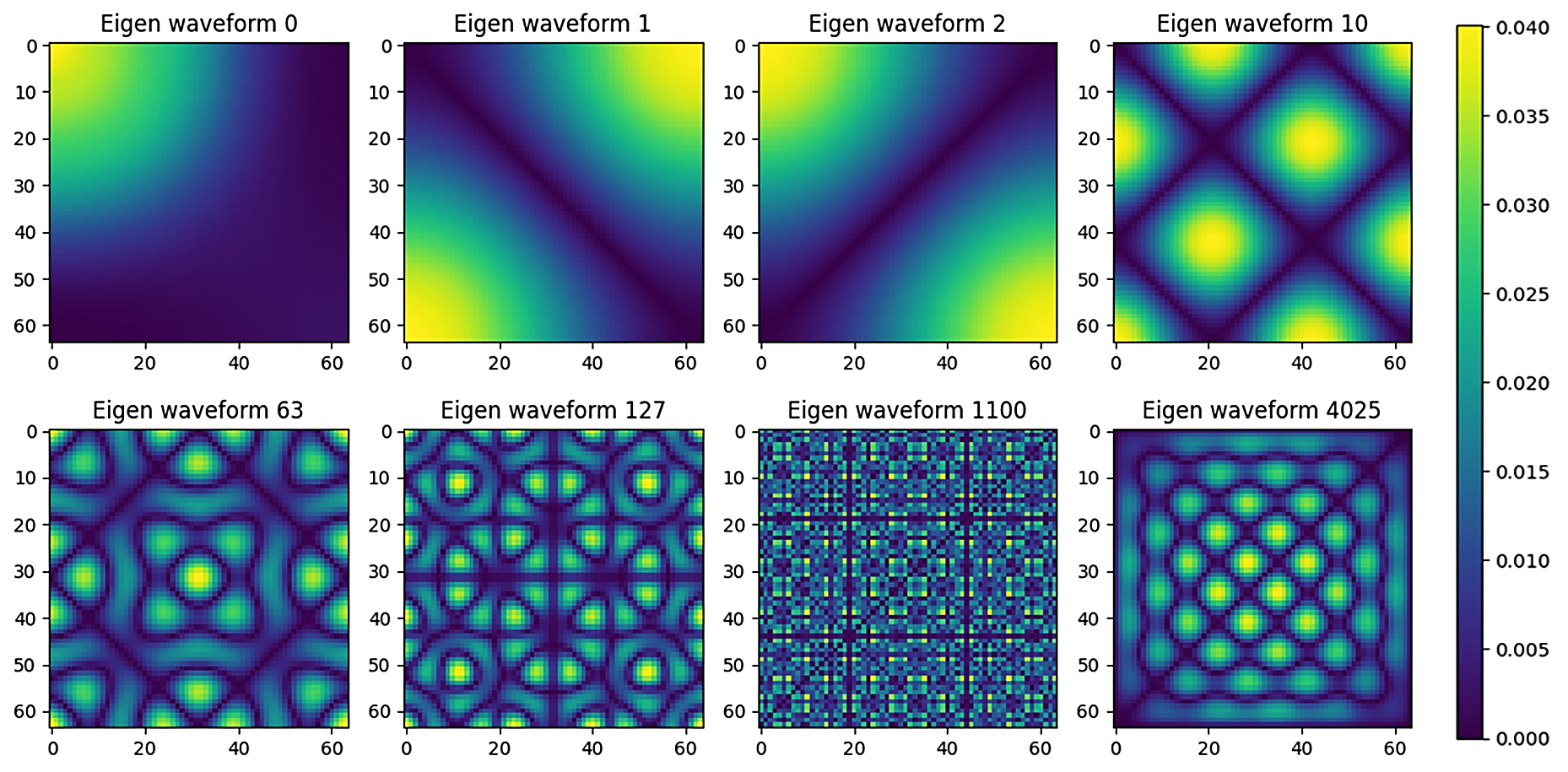}}
\captionsetup{font=normal}
\caption{Eigen waveforms illustrating select low and high frequency modes of the defined graph structure}
\label{Fig: Figure 3}
\end{figure*}

Moving beyond these foundational eigenvectors, the subsequent high-frequency modes unveil localized variations within the image grid. These high-frequency eigenvectors delve into fine-grained features, highlighting intricate patterns and details that might be imperceptible in lower-frequency modes. As eigenvalues ascend, each associated eigenvector offers a more refined perspective on the spatial relationships and structural intricacies within the image. Figure ~\ref{Fig: Figure 3} provides a visual representation of few select eigen vectors having low and high frequency waveforms.

\subsection{Graph fourier transform}

The graph Fourier transform (GFT) is a mathematical tool applied to signals defined on graphs. For a given graph Laplacian operator $L = U\Lambda U^T$, where $U$ is the matrix of eigenvectors and $\Lambda$ is the diagonal matrix of corresponding eigenvalues, the GFT of a signal $X=(x_1,x_2,\ldots,x_N)$ is defined as

\begin{equation}
    S = U^T X, \quad s_k = \langle X, u_k \rangle = \sum_{i=1}^{N} x_i \cdot u_k(i-1)
\end{equation}

\noindent where spectral coefficient \(s_k = \langle X, u_k \rangle\) is the inner product between the signal \(X\) and eigenvector \(u_k\), representing the projection of the signal onto the \(k\)\textsuperscript{th} graph frequency mode.

Anomalous patterns or defects within an image often manifest as distinct localized variations, stemming from diverse physical mechanisms encountered throughout various stages of the manufacturing process. The spectral coefficients \(S=\{s_0, s_1, s_3, \ldots, s_{N-1}\}\) derived by projecting the signal onto the eigenvectors not only accentuate the presence of defects in the graph frequency domain but also provide a distinctive signature for precise defect localization.

Leveraging the orthogonality property of the eigenvector matrix (\(U \cdot U^T = I\)), the original image can be precisely reconstructed using \(X = US\). In sections 3 and 4, we employ a 1D-CNN binary classifier using spectral coefficients from an image dataset characterized by diverse background textures, including images with defects and non-defect scenarios. Employing SHAP explanations, we identify important spectral coefficients \(S^{*}\) for each image. The localization of defects becomes visually discernible through the reconstruction process \(X^* = US^*\), capitalizing on the essential spectral coefficients that encapsulate distinct signatures of the defects.

\section{1D Convolutional neural network classifier}

\begin{figure}[ht]
    \centering
    \includegraphics[width=0.90\linewidth]{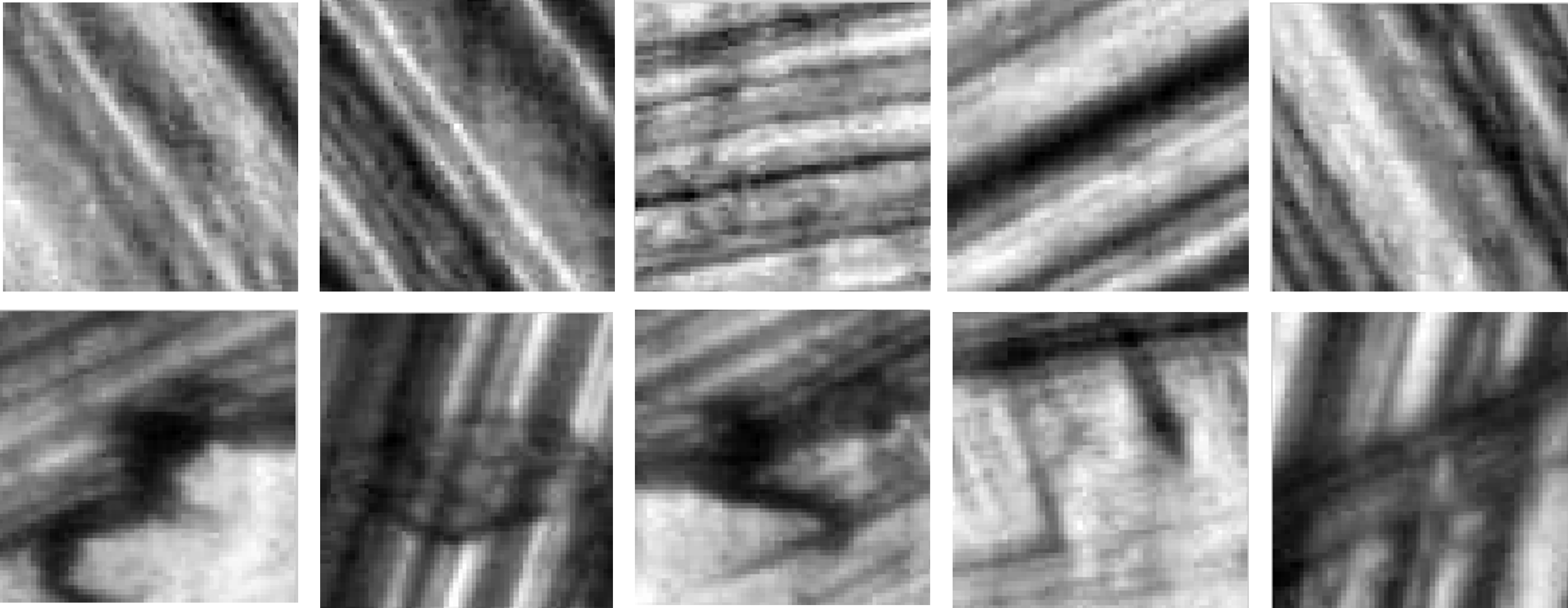}
    \captionsetup{font=normal}
    \caption{Top row containing images with diverse background textures and no defects, and bottom row showcasing images with defects.}
    \label{Fig:Figure 4}
\end{figure}

We construct a binary classifier using 1D-CNN, taking the spectral coefficients as inputs from an image dataset characterized by diverse background textures, including images with defects, and non-defect scenarios. We utilized a dataset of 256 $\times$ 256 surface images obtained from the turning process, generously provided by our collaborator Marposs-STIL. This dataset comprises 228 instances of defects and 800 non-defective images. To prepare for the subsequent classification task, we extracted 1000 defect and non-defect patches, each measuring 64 $\times$ 64, from the original dataset, as illustrated in Figure \ref{Fig:Figure 4}. Employing set rotation and flipping for augmentation, we expanded the dataset to include 6000 defect and 6000 non-defect patches. Following this, Graph Fourier Transform (GFT) features were extracted from each patch, resulting in a final dataset sized 12000 $\times$ 4096. To assess the effectiveness of our methodology, an 80-20 split was employed, allocating 80\% of the dataset for training a 1D Convolutional Neural Network (CNN) model on the GFT features. This model was purposefully designed to classify patches into defect and non-defect categories, leveraging GFT features as the basis for our exploration into surface image defect localization.

\begin{figure}[ht]
\centering
\includegraphics[width=1\linewidth]{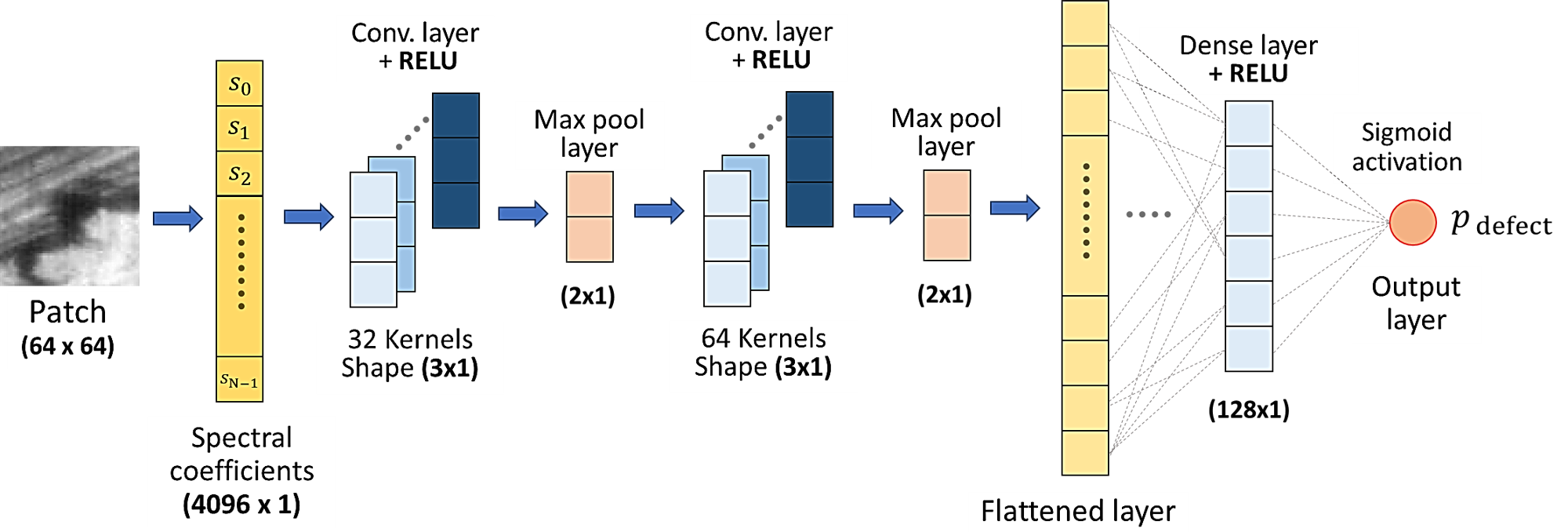}
\captionsetup{font=normal}
\caption{1D-CNN model architecture}
\label{Fig:Figure 5}
\end{figure}

\begin{figure}[ht]
\centering
\includegraphics[width=0.9\linewidth]{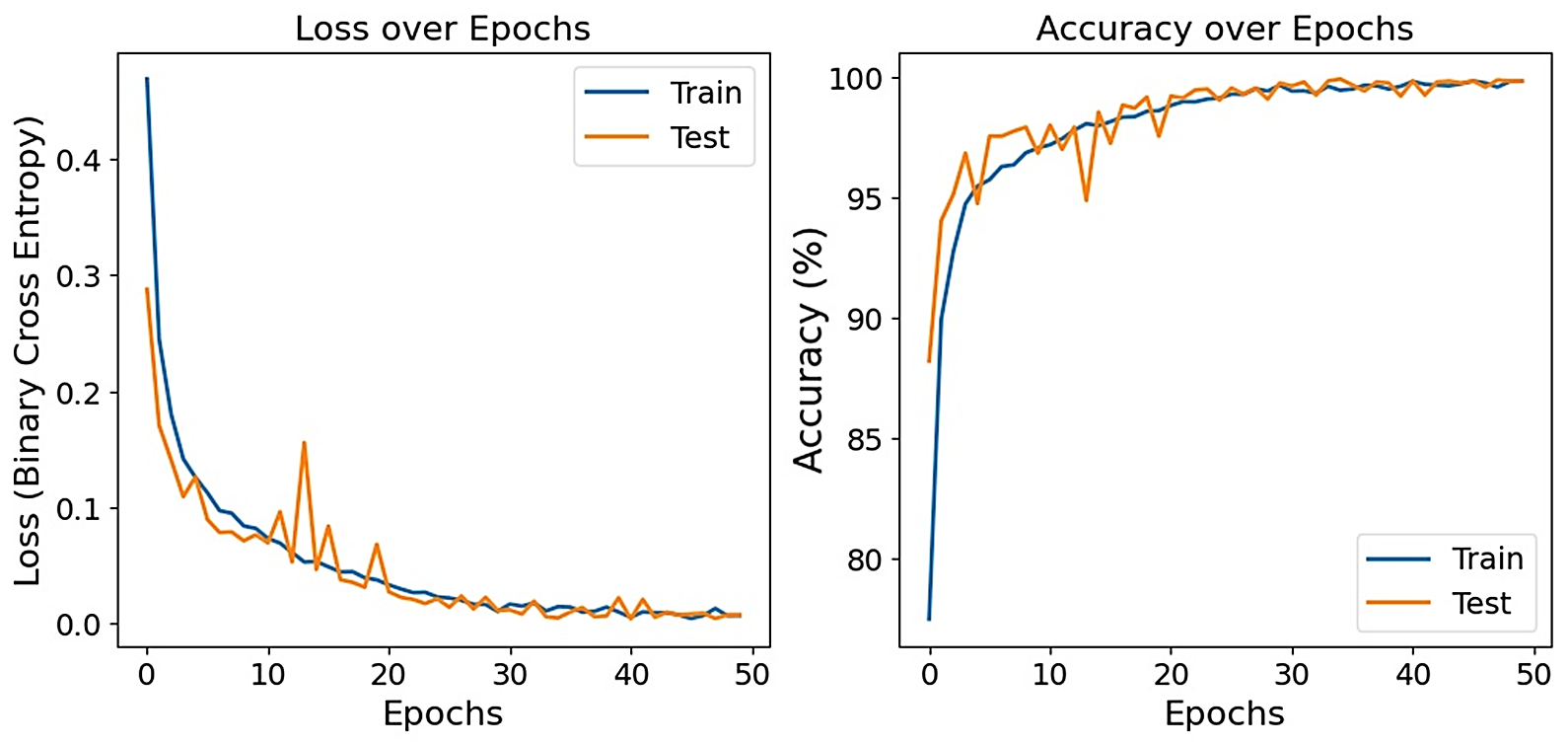}
\captionsetup{font=normal}
\caption{Performance of 1D-CNN binary classifier}
\label{Fig: Figure 6}
\end{figure}

The proposed model architecture for defect classification, illustrated in Figure \ref{Fig:Figure 5}, involves a sequence of two 1D Convolution layers with ReLU activation functions, utilizing 32 and 64 kernels and a stride of 3. Following the convolutional layers, the output is flattened and subjected to a dropout layer. Subsequently, two fully connected layers are integrated, featuring hidden neurons arranged as (128, 2), accompanied by ReLU and Sigmoid activation functions, respectively. The model is compiled using categorical cross-entropy as the chosen loss function, the Adam optimizer (learning rate = 0.001), and accuracy as the primary evaluation metric. The training regimen spans 50 epochs with a batch size of 64, utilizing labeled image vectors for both training and testing phases. This architectural configuration, along with the specified training parameters, is strategically designed to enhance the model's proficiency in accurately classifying defects and non-defects within the surface image dataset. Figure \ref{Fig: Figure 6} shows that the model achieves a training accuracy of 99.94\% and a testing accuracy of 99.91\%, underscoring its robust performance in accurately classifying defects and non-defects within the surface image dataset.

To validate the significance of spectral coefficients in classification, we employed a Random Forest Classifier with these coefficients as features, achieving notable training and testing accuracies of 99.89\% and 96.87\%, respectively. This underscores the discriminative power of spectral coefficients in defect identification. Subsequently, our focus shifts to SHAP explanations from the trained 1D-CNN classifier in the upcoming section. This analysis aims to pinpoint essential spectral coefficients crucial for defect localization, providing key insights into the model's decision-making process.

\section{Results and discussion}

After successfully classifying an image as defect or non-defect using the trained 1D CNN model \(f\), we aim to identify the GFT spectral coefficients that exert the most influence on the model's output. These dominant coefficients provide insights into the defect localization process, enabling us to move beyond binary classification and pinpoint the specific spectral coefficients associated with defects. 

Shapley value is a game theory-based metric widely used for quantifying the contribution of each player in a cooperative game. In our context, Shapley values measure the importance of each spectral coefficient in determining whether an image contains a defect. Specifically, the contribution of a coefficient $s_i$ to the model output for an input $X$ is determined by the difference between outputs when including and excluding coefficient $s_i$, expressed as $f({X}_{R\cup\{s_i\}}) - f({X}_{R})$. Here, $R$ is a subset of $S\setminus\{s_i\}$ (the entire set of coefficients $S$ excluding the coefficient $s_i$). ${X}_{R}$ represents a modified input that only contains coefficients in $R$ and takes the same coefficient values as ${X}$ does. Overall, the marginal contribution (or Shapley value) of coefficient $s_i$ regarding the input $X$ (denoted as $\phi_{s_i}^{X}$) is computed as a weighted average of the output differences over all possible coefficient subsets, that is 

\begin{equation} \label{eqshapley}
    \phi_{s_i}^{X} = \sum_{R\subseteq S\setminus\{s_i\}} \frac{|R|!(|S|-|R|-1)!}{|S|!} [f({X}_{R\cup\{s_i\}}) - f({X}_{R})]
\end{equation}
wherein $|\cdot|$ is the size of a set. $\phi_{s_i}^{\bf x}$ satisfies 

\begin{equation} \label{eqshapleysum}
    \phi_{0}+\sum_{i=1}^{|S|} \phi_{s_i}^{X} = f({X})
\end{equation}

where $f({X})$ is the original model output on ${X}$. $\phi_{0}$ denotes the model output when all spectral coefficients are zero, i.e., $\phi_{0}$ is a constant based on the model's historical training data regardless of ${X}$. Based on Eq. (\ref{eqshapleysum}), a positive Shapley value for a particular coefficient of a defect image indicates an increased likelihood of the image being classified as defective, and vice versa. 

However, the computation complexity of Eq.(\ref{eqshapley}) can be extrememly large when considering all possible coefficient subsets \cite{shap}. Therefore, SHAP, particularly the DeepSHAP algorithm in our case, approximates the Shapley value using selected representative subsets. DeepSHAP also approximates $f({X}_{R\cup\{s_i\}}) - f({X}_{R})$ for a given coefficient subset $R$ by leveraging the gradient information obtained through backpropagation \cite{shap}. The resulting estimate of the Shapley value for coefficient $s_i$ is referred to as the SHAP value, denoted by $\hat{\phi}_{s_i}^{X}$. When applied to individual images, SHAP unveils important spectral coefficients to each image, contributing to effective defect localization. In addition, SHAP also provides a global inference perspective, which considers the collective impact of coefficients across the entire dataset. Specifically, after obtaining the SHAP value of coefficient $s_i$ for every image ${X}^{(j)}$, $j= 1,2,...,M$, the algorithm aggregate the SHAP values over all $M$ images to derive the global contribution of coefficient $s_i$, denoted by $G_{s_i}$, as 

\begin{equation}\label{eqglobal}
G_{s_i} = \sum_{j=1}^{M} |\hat{\phi}_{s_i}^{{X}^{(j)}}|,\qquad  s_i \in S
\end{equation}

\begin{figure}[ht]
\centering
\includegraphics[width=0.67\linewidth]{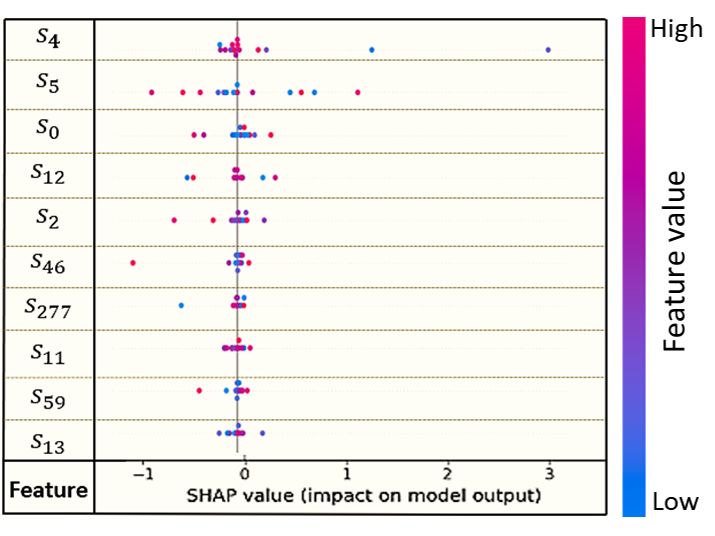}
\captionsetup{font=normal}
\caption{Summary plot of SHAP values for a subset of correctly classified 15 defect images}
\label{[Fig: Figure 7}
\end{figure}

\begin{figure}[ht]
\centering
\includegraphics[width=0.67\linewidth]{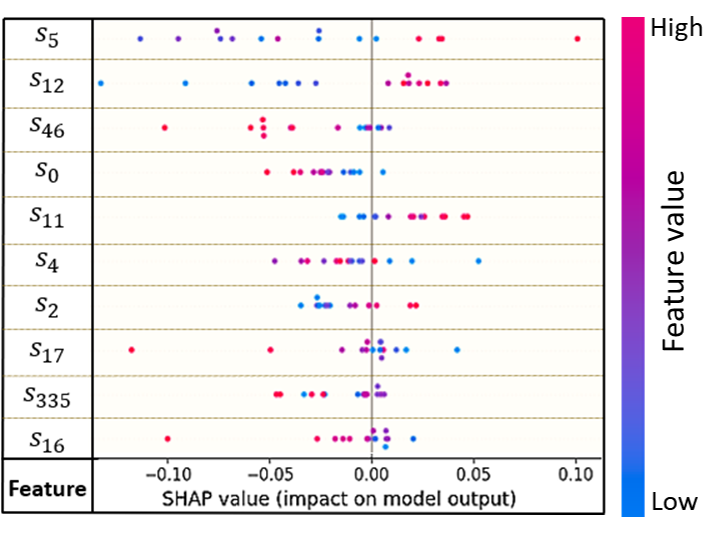}
\captionsetup{font=normal}
\caption{Summary plot of SHAP values for a subset of correctly classified 15 non-defect images}
\label{[Fig: Figure 8}
\end{figure}

Two curated data subsets, each comprising 15 images, encompass instances of defects and non-defects. SHAP values for each of the 15 images are subsequently derived using SHAP DeepExplainer, trained on 500 images from the overall training dataset, which includes both classes. The outcomes demonstrate a pronounced discriminative capability between images with defects and those without, particularly emphasizing the importance of low-frequency spectral coefficients. Cumulative summaries of the top 10 important spectral coefficients for classifying each data class are visually presented in Figures \ref{[Fig: Figure 7} and \ref{[Fig: Figure 8}, offering insights into the key features influencing the classification process.

\begin{figure}[ht] 
\centerline{\includegraphics[width=0.91\linewidth]{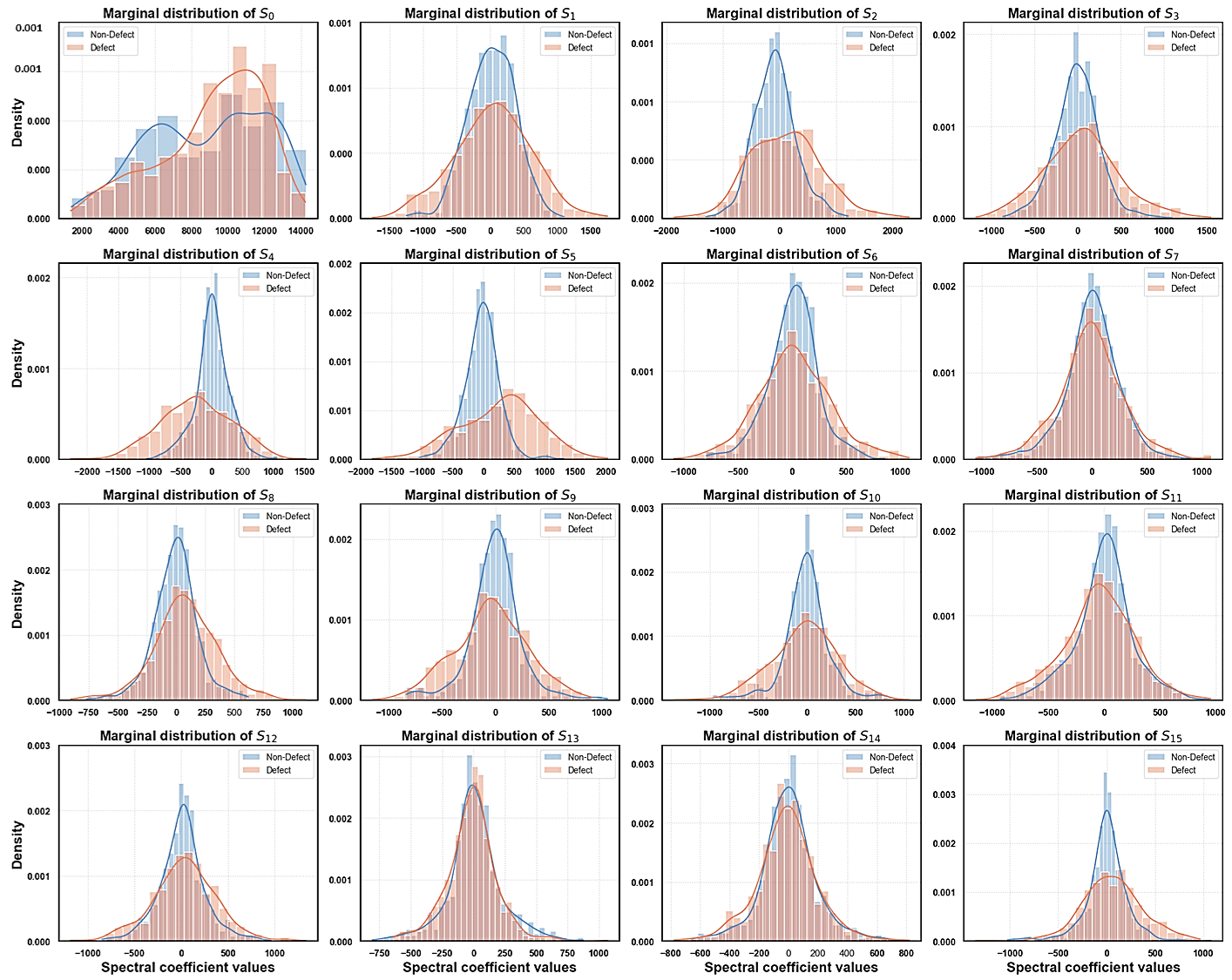}}
\captionsetup{font=normal}
\caption{Marginal distribution of low-frequency spectral coefficients for defct and non-defect images}
\label{Fig: Figure 9}
\end{figure}

\begin{figure}[ht]
\centerline{\includegraphics[width=0.92\linewidth]{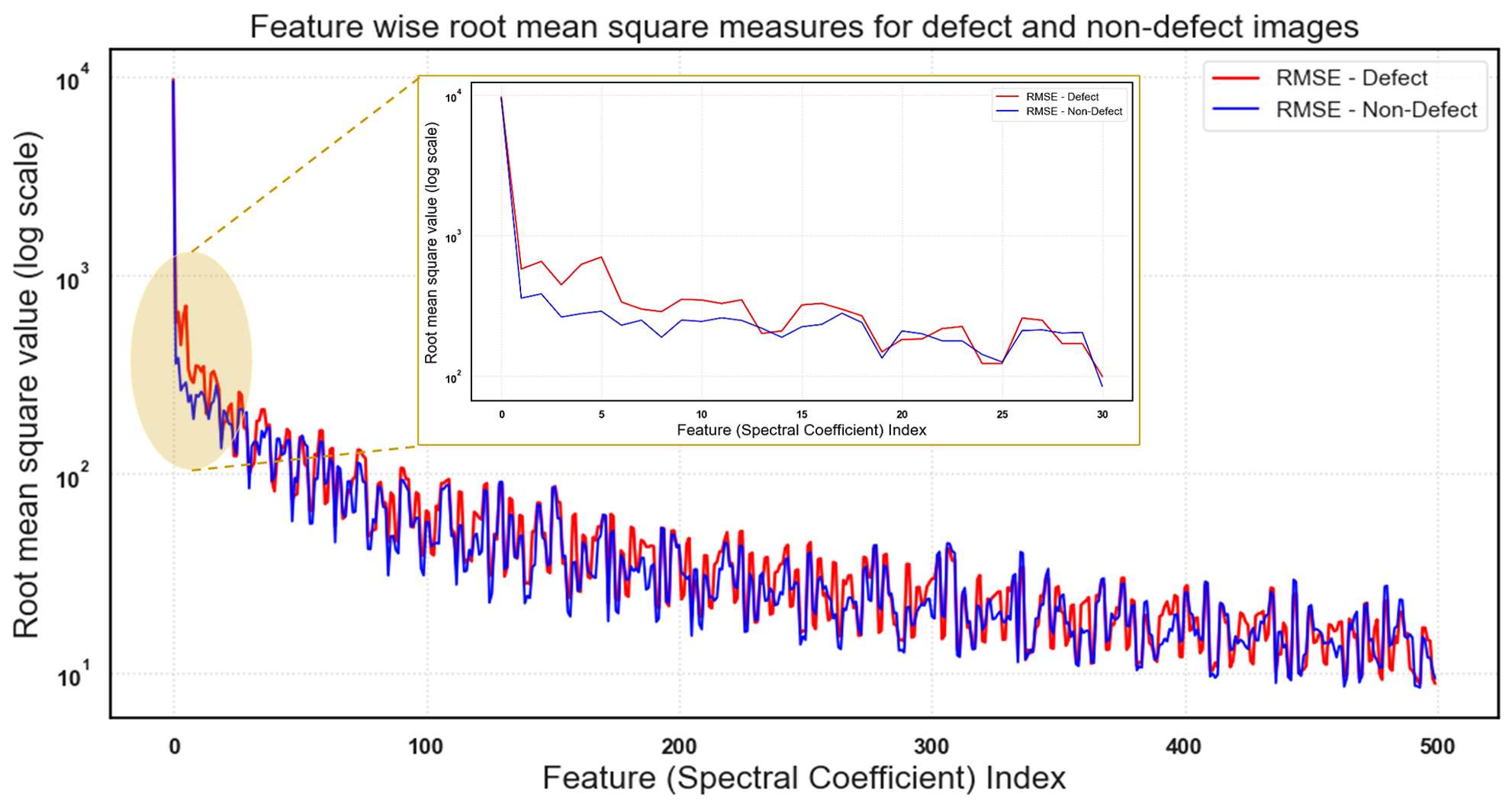}}
\captionsetup{font=normal}
\caption{Root mean square measures of the first 500 low-frequency spectral coefficients for defect and non-defect images}
\label{Fig: Figure 10}
\end{figure}

Subsequent to this analysis, we examine the marginal distribution of the first 16 low-frequency spectral coefficients, denoted as ${s_0}$ to ${s_{15}}$, across the entire dataset for each class. As emphasized by SHAP, a noteworthy shift in median values is observed, particularly for ${s_4}$, ${s_5}$, and ${s_2}$. Additionally, Fig \ref{Fig: Figure 9} illustrates the distinct dissimilarity in the marginal distribution of spectral coefficient ${s_0}$ between the two classes. The defects examined in this study demonstrate a tendency to disturb the low-frequency eigen waveforms in contrast to non-defect images with distinctive background textures as observed in Figure \ref{Fig: Figure 9}. The pronounced discriminative capability between defect and non-defect images, as clarified by SHAP through low-frequency spectral coefficients, receives additional validation through the Root Mean Square plot of the spectral coefficients across the entire test dataset (Fig \ref{Fig: Figure 10}).

\begin{figure*}[t]
\centerline{\includegraphics[width=1.01\linewidth]{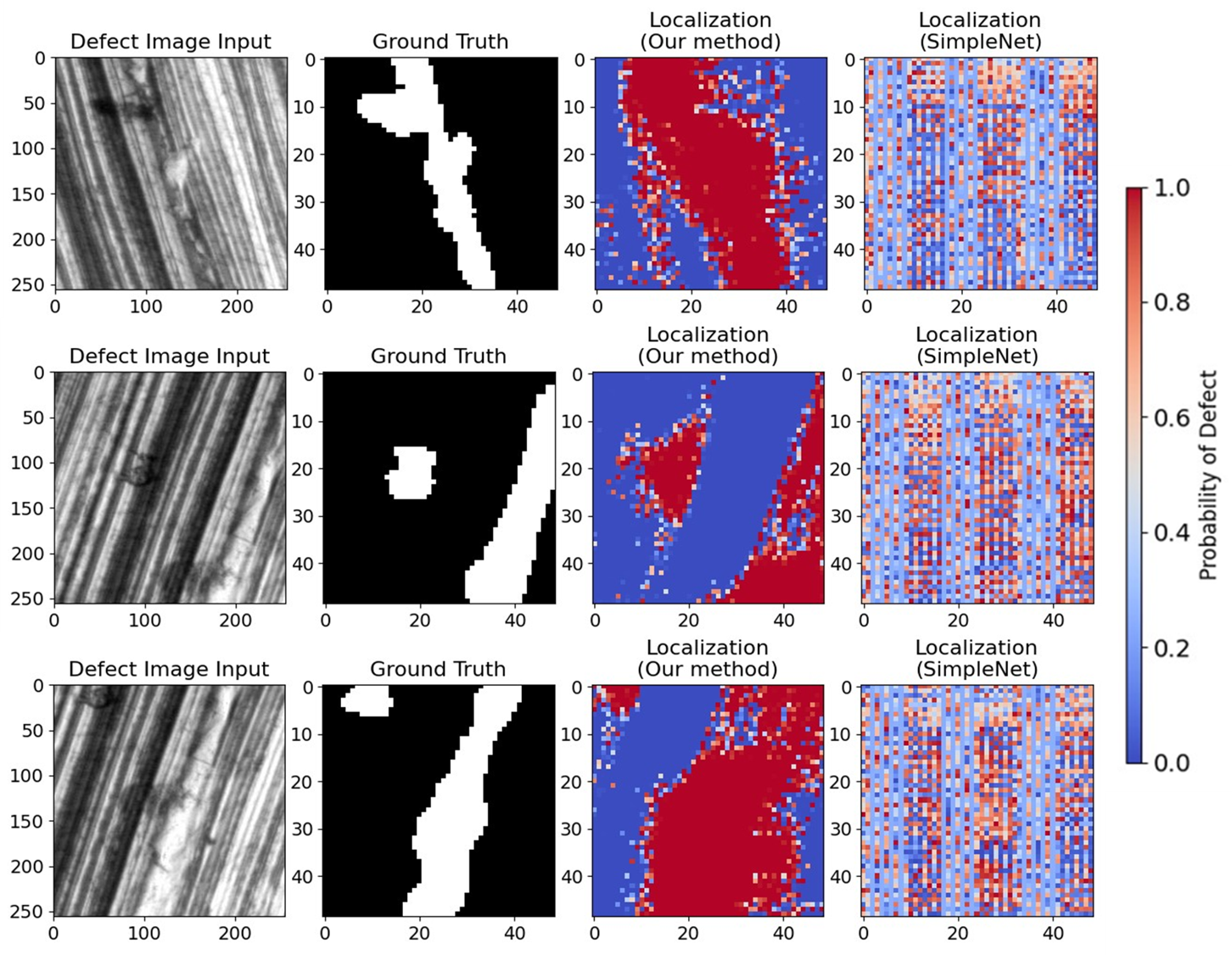}}
\captionsetup{font=normal}
\caption{Defect localization outputs: Defect Input Image, Ground truth, localization (Our method), and localization (SimpleNet)}
\label{Fig: Figure 11}
\end{figure*}

\begin{figure}[ht]
\centerline{\includegraphics[width=1.05\linewidth]{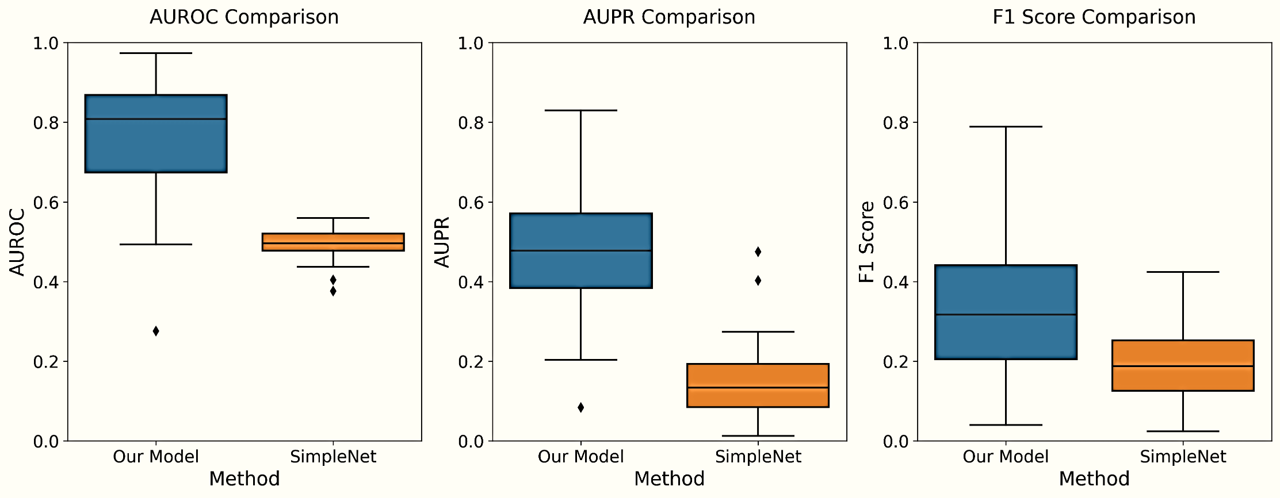}}
\captionsetup{font=normal}
\caption{Defect localization performance: Our Model vs SimpleNet}
\label{Fig: Figure 12}
\end{figure}

Defect localization on the 256 $\times$ 256 surface image is conducted by classifying each 64x64 patch with a stride of 8 across the entire image, leveraging the capabilities of the trained 1D CNN model. The outcome is a sparse defect localization output of dimensions 49x49. The defect localization results from our model are visually presented in the third column of Figure \ref{Fig: Figure 11}, providing a representation of the spatial localization of identified defects across the surface image.

As discussed in Section 1 and illustrated in Figure \ref{Fig:Figure 1}, the surfaces derived from various manufacturing processes exhibit complex background patterns, complicating the development and training of algorithms tailored for defect localization unique to the process. The proposed method revolves around the core idea of capturing these complexities through the utilization of well-established GFT features. Our defect localization approach is then assessed through a comparative analysis with SimpleNet \cite{liu2023simplenet}, a widely recognized state-of-the-art methodology which stands out as a simple yet application-friendly network. This comparative study highlights the challenge of formulating effective and customized methods for defect localization on intricate machining textures. On the MVTec AD benchmark, SimpleNet \cite{liu2023simplenet} demonstrates notable performance, achieving an anomaly detection AUROC of 99.6\% and reducing errors by 55.5\% compared to the next best-performing model. The SimpleNet is trained using hyperparameters consistent with those employed by its original authors for their study on the MVTEC-AD dataset. Figure \ref{Fig: Figure 11} visually compares the defect localization outputs from both models, juxtaposed with the ground truth. Further quantitative insights are presented in Figure \ref{Fig: Figure 12}, featuring boxplots comparing the models' performance using AUROC, AUPR, and F1 score metrics. Our observations reveal that our model excels in the sparse localization and detection of defects within surface images.

\section{Conclusion and Future Work}

This work is one of the first investigations of graph Fourier analysis for detection and localization of defects and such artifacts from images with highly textured background. The approach harnesses the power of graph representations to encapsulate intricate dynamics inherent in high-dimensional data, specifically image pixels, while simultaneously preserving vital locality properties within a lower-dimensional space, here, the graph. Through the application of Graph Fourier Transforms, the extraction of essential spectral coefficients from the image facilitates localization of defects and artifacts within images characterized by highly textured backgrounds.

A convolutional neural network model (1D-CNN) trained on the graph Fourier spectral coefficients was able to classify, with classification accuracy exceeding 99\%, whether the image contains a defect. The CNN classifier was able to exceed the performance of perhaps the most contemporary image-based defect (artifact) detection method (called SimpleNet) by 70\% in terms of the area under the receiver operating characteristic (AUROC), and F1 score. The subsequent application of SHAP, an explainable AI (XAI) method enabled the identification of the key components of the graph Fourier spectrum that inform the defects, including its location and other morphological characteristics. The SHAP analysis suggests that the eigen waveforms 4 and 5 (that capture the lower frequency components) are most sensitive to the presence of the defects. 

The current research assumes a constant graph structure to derive spectral coefficients for each patch. A promising extension involves formulating a dataset-specific graph structure through the utilization of Graph Neural Networks (GNNs). By employing GNNs, we can dynamically construct a unique graph topology tailored to the dataset characteristics. This innovative approach would enable the derivation of spectral coefficients for patches, potentially enhancing the efficiency of defect localization in surface images. This extension holds the potential to further refine the model's adaptability to diverse datasets, contributing to more effective and nuanced defect identification methodologies in manufacturing processes.

\label{main}

\section*{Acknowledgements}
This work was supported by Rockwell International Professorship and U.S. National Science Foundation under Grant NSF-ECCS-1953694. The authors would also like to express their sincere gratitude to MARPOSS-STIL for generously providing all the image data. 

\bibliographystyle{unsrt}
\bibliography{refs.bib}

\end{document}